\documentclass{PoS}

\title{$F_{2}^{b\bar{b}}$ measurement at ZEUS}

\ShortTitle{$F_{2}^{b\bar{b}}$ at ZEUS}

\author{\speaker{Philipp Roloff}$^{ab}$\thanks{On behalf of the ZEUS Collaboration.} \\
        \llap{$^{a}$}Deutsches Elektronen-Synchrotron DESY, Notkestr. 85, D-22607 Hamburg, Germany \\
        \llap{$^{b}$}University of Hamburg, Institute of Experimental Physics, Luruper Chaussee 149, D-22761 Hamburg, Germany \\
        E-mail: \email{philipp.roloff@desy.de}}

\abstract{Two recent measurements of beauty production in deep inelastic scattering based on data collected by the ZEUS detector are summarised. In the first one, the beauty fraction in the data was obtained from events with a muon and a jet. In the second one, beauty cross sections were measured using the decay length significance and mass of inclusive secondary vertices. Differential cross sections are presented and compared to QCD predictions. The beauty contribution to the inclusive proton structure function $F_{2}$ was extracted for the jet+muon measurement and is compared to previous measurements and theoretical predictions.}

\FullConference{XVIII International Workshop on Deep-Inelastic Scattering and Related Subjects, DIS 2010\\
		April 19-23, 2010\\
		Firenze, Italy}

\begin{document}

\section{Introduction}

Beauty production in the deep inelastic scattering (DIS) regime (squared four-momentum exchange at the electron vertex $Q^{2} \gtrsim$ a few GeV$^{2}$) is dominated by the boson-gluon fusion (BGF) process, where the virtual photon interacts with a gluon from the proton. Hence this process is directly sensitive to the gluon content of the proton. The contribution of beauty production to the inclusive proton structure function $F_{2}$ can be extracted from double differential cross sections in the Bj{\o}rken scaling variable $x$ and $Q^{2}$.

Mass effects are relevant in a large part of the kinematic region for beauty production in DIS accessible at HERA. The precise measurement of beauty production could thus help to distinguish between different theoretical approaches to include mass effects in perturbative QCD calculations.

\section{Measurement of beauty production using events with a muon and a jet}
\label{sec:muon_and_jet}

The production of beauty quarks with events in which a muon and a jet are observed in the final state has been measured with the ZEUS detector at HERA using an integrated luminosity of $114.1 \pm 2.3$~pb$^{-1}$~\cite{ref:muon_and_jet}. The analysed data were collected in the years from 1996 to 2000.

The reaction $ep \rightarrow e' b\bar{b} X \rightarrow e' {\rm jet} X'$ was measured in the kinematic region defined by: $Q^{2}>2$~GeV$^{2}$, $0.05 < y < 0.7$, and at least one jet with $E_{T}^{{\rm jet}}>5$~GeV and $-2 < \eta^{{\rm jet}} < 2.5$ including a muon of $p_{T}^{\mu} > 1.5$~GeV and $\eta^{\mu} > -1.6$ inside a cone $\Delta R<0.7$ to the jet axis. Jets were obtained using the $k_{T}$ cluster algorithm at the hadron level in its massive mode with the $E_{T}$-recombination scheme. Weakly decaying B-hadrons were treated as stable particles and were decayed (e.g. to a muon) only after the application of the jet algorithm.

\begin{figure}[h]
\centering
\includegraphics[width=0.5\textwidth]{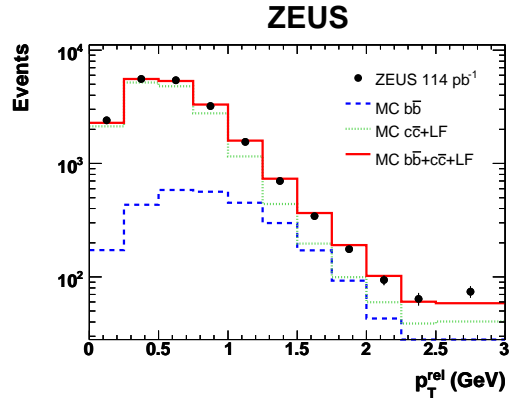}
\caption{The measured $p_{T}^{{\rm rel}}$-distribution in the data compared to the Monte Carlo predictions for beauty and background scaled to the fit result.}
\label{fig:ptrel}
\end{figure}

The beauty signal was extracted using the distribution of the transverse momentum of the muon with respect to the momentum vector of the associated jet, $p_{T}^{{\rm rel}}$. The fraction of beauty and background events in the data sample was obtained from a two-component fit to the shape of the measured $p_{T}^{{\rm rel}}$ distribution. The $p_{T}^{{\rm rel}}$ distribution for beauty was taken from the RAPGAP Monte Carlo (MC) program~\cite{ref:rapgap}. The corresponding distribution for the background was obtained from the sum of the light flavour (LF) distribution taken from the ARIADNE~\cite{ref:ariadne} model and the charm distribution taken from RAPGAP weighted according to the cross sections predicted by the two MC programs. The measured $p_{T}^{{\rm rel}}$-distribution and the MC distributions for beauty and background scaled to the fit result are shown in Fig.~\ref{fig:ptrel}.

\begin{figure}[h]
\centering
\includegraphics[width=0.5\textwidth]{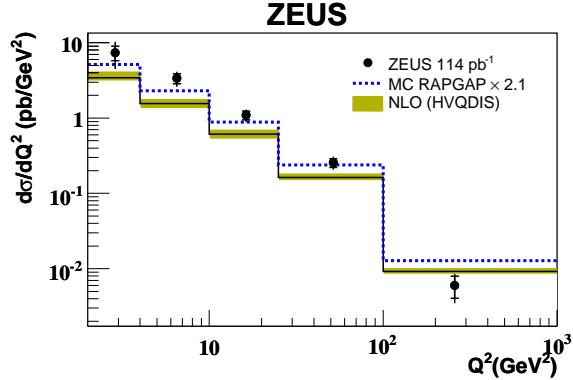}
\caption{Differential beauty cross section as a function of $Q^{2}$ compared to the RAPGAP Monte Carlo normalised to the data and compared to the HVQDIS NLO QCD prediction.}
\label{fig:q2}
\end{figure}

Figure~\ref{fig:q2} shows the measured differential cross section as a function of $Q^{2}$ compared to the HVQDIS~\cite{ref:hvqdis} NLO QCD calculation and the RAPGAP MC scaled to the data. Both predictions describe the shape of the data reasonably well. The region at $Q^{2} \lesssim 5$~GeV$^{2}$ can only be measured using data taken during the HERA I period and is thus unique to this measurement.

\begin{figure}[t]
\centering
\includegraphics[width=0.6\textwidth]{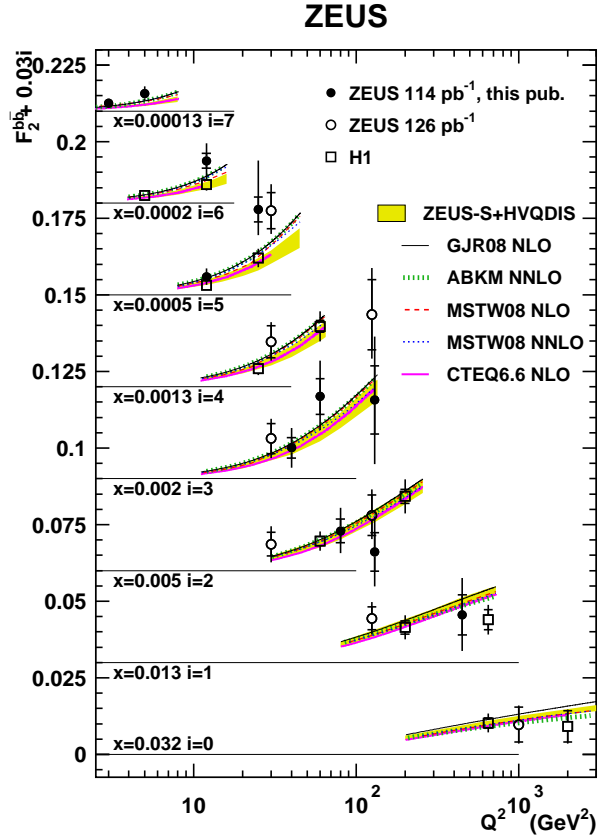}
\caption{Measured $F_{2}^{b\bar{b}}$ as a function of $Q^{2}$ (filled circles). Results from previous measurements (open symbols) and from different QCD predictions (lines and band) are also shown.}
\label{fig:f2b}
\end{figure}

The double-differential cross section for $b\bar{b}$-production in DIS can be expressed as:
\begin{equation}
\frac{d^{2} \sigma^{b \bar{b}}(x,Q^{2})}{dx dQ^{2}} = \frac{2 \pi \alpha^{2}}{Q^{4} x} \left\{ [1 + (1 - y)^{2}] F^{b \bar{b}}(x,Q^{2}) - y^{2} F^{b \bar{b}}_{L}(x,Q^{2}) \right\}.
\end{equation}
The contribution from $F_{L}^{b\bar{b}}$ is small for the measured $Q^{2}$ and $x$ ranges and was neglected. To extract $F_{2}^{b\bar{b}}$ from the measured double-differential cross sections in bins of $x$ and $Q^{2}$, an extrapolation to the full kinematic phase space was performed using HVQDIS:
\begin{equation}
F_{2,{\rm meas}}^{b\bar{b}}(x_{i},Q_{i}^{2}) = \frac{\sigma_{{\rm meas},i}}{\sigma_{{\rm HVQDIS},i}} \times F_{2,{\rm HVQDIS}}^{b\bar{b}}(x_{i},Q^{2}_{i}).
\end{equation}
The prediction for $F_{2}^{b\bar{b}}$ from HVQDIS, $F_{2,{\rm HVQDIS}}^{b\bar{b}}$, is multiplied by the ratio of the measured, $\sigma_{{\rm meas},i}$, to the predicted, $\sigma_{{\rm HVQDIS},i}$, visible cross sections in a given bin $i$. The results of the extraction are shown in Fig.~\ref{fig:f2b} and compared to a previous ZEUS measurement~\cite{ref:zeus_2005} focusing on the higher $Q^{2}$ region and H1 measurements~\cite{ref:h1} based on lifetime information. The data are all compatible within uncertainties. At low $x$, the ZEUS data have a tendency to lie slightly above the H1 measurements. The predictions from different theoretical approaches~\cite{ref:predictions} agree fairly well with each other. The HVQDIS predictions are somewhat lower than the ZEUS data at low $Q^{2}$ and $x$ while at higher $Q^{2}$ the data are described by all predictions.

\section{Measurement of beauty production using inclusive secondary vertices}

\begin{figure}[ht]
\centering
\includegraphics[width=0.5\textwidth]{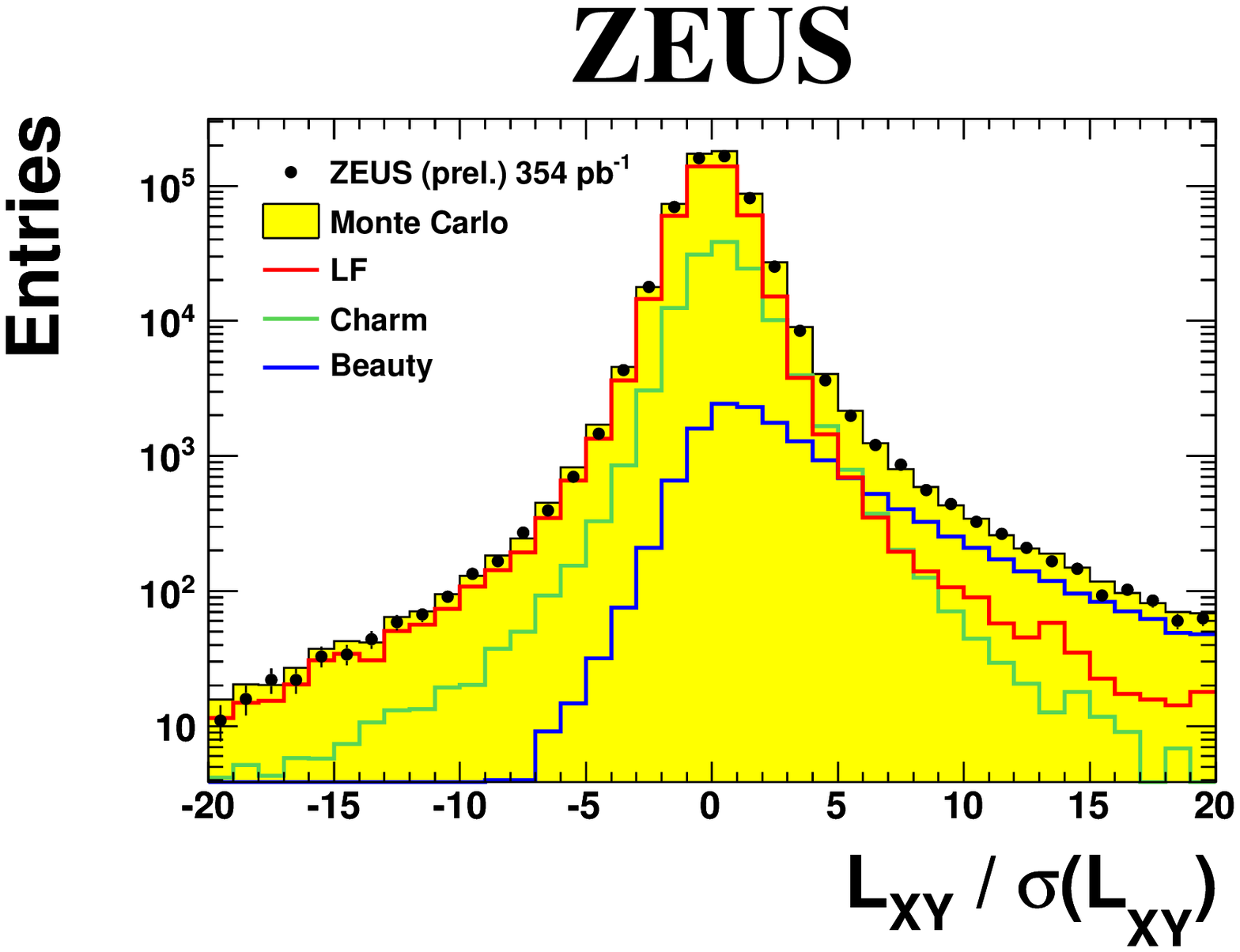}
\caption{Decay length significance, $S$, for the data and the different Monte Carlo subsamples in the region $2 < M_{{\rm VTX}} < 6$~GeV.}
\label{fig:significance}
\end{figure}

Beauty quark production has also been measured using the decay length significance and mass of inclusive secondary vertices using an integrated luminosity of $354 \pm 9$~pb$^{-1}$ recorded in the years from 2004 to 2007. The kinematic range of the measurement is given by: $5 < Q^{2} < 1000$~GeV$^{2}$, $0.02 < y < 0.7$, $E_{T}^{{\rm jet}} > 5$~GeV and $-1.6 < \eta^{{\rm jet}} < 2.2$. The jets were reconstructed using the same algorithm as described in Section~\ref{sec:muon_and_jet}.

To reconstruct B-hadron decay vertices, tracks well reconstructed by the central tracking detector (CTD) and the micro vertex detector (MVD) with a transverse momentum of $p_{T} > 500$~MeV were considered. These tracks were associated to a jet if they fulfilled the cut $\Delta R < 1$ with $\Delta R = \sqrt{\Delta\eta^{2} + \Delta\phi^{2}}$. A secondary vertex was fitted for any selected jet if at least two tracks were associated to the jet. The weights of tracks not well fitting to the secondary vertex were reduced by a deterministic annealing filter.

The beauty content in the selected sample was determined using the decay length significance $S = L_{XY} / \sigma(L_{XY})$. $S$ is defined as the distance between the primary interaction vertex (beam spot) and the secondary vertex, $L$, divided by its uncertainty and projected onto the jet axis in the $XY$-plane. The significance distribution observed in the data and for beauty and charm MC samples generated using RAPGAP and a LF distribution predicted by the ARIADNE generator are shown in Fig.~\ref{fig:significance}.

\begin{figure}[b]
\centering
\includegraphics[width=0.6\textwidth]{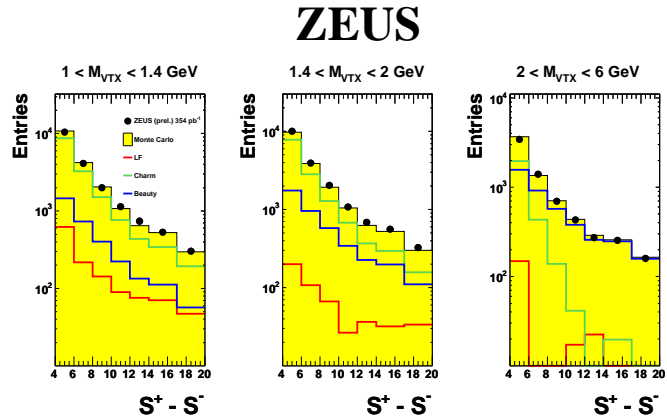}
\caption{Mirrored decay length in three bins of the vertex mass.}
\label{fig:mirrsig}
\end{figure}

The left side of the decay length significance distribution ($S^{-}$, $S < 0$) was mirrored onto and subtracted from the right side ($S^{+}$, $S > 0$). Only the region with $|S| > 4$ was used to enhance the beauty fraction. Finally, the mirrored decay length significance was divided into three different mass bins to provide an almost pure beauty sample in the region $2 < M_{{\rm VTX}} < 6$~GeV (see Fig.~\ref{fig:mirrsig}). The other two bins are dominated by charm.

\begin{figure}[t]
\begin{minipage}[t]{0.24\textwidth}
\centering
\includegraphics[width=0.9\textwidth]{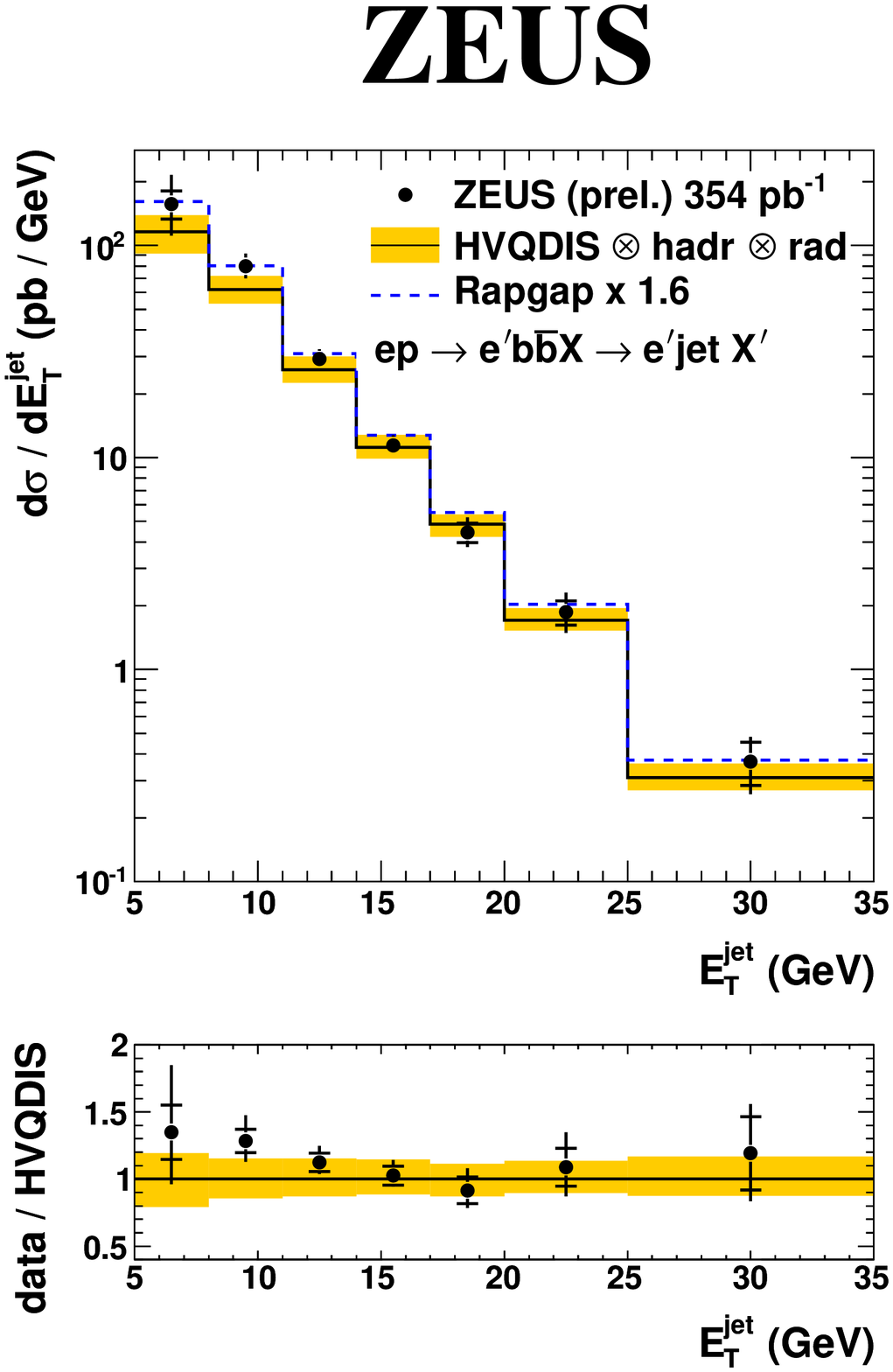}
\end{minipage}
\begin{minipage}[t]{0.24\textwidth}
\centering
\includegraphics[width=0.9\textwidth]{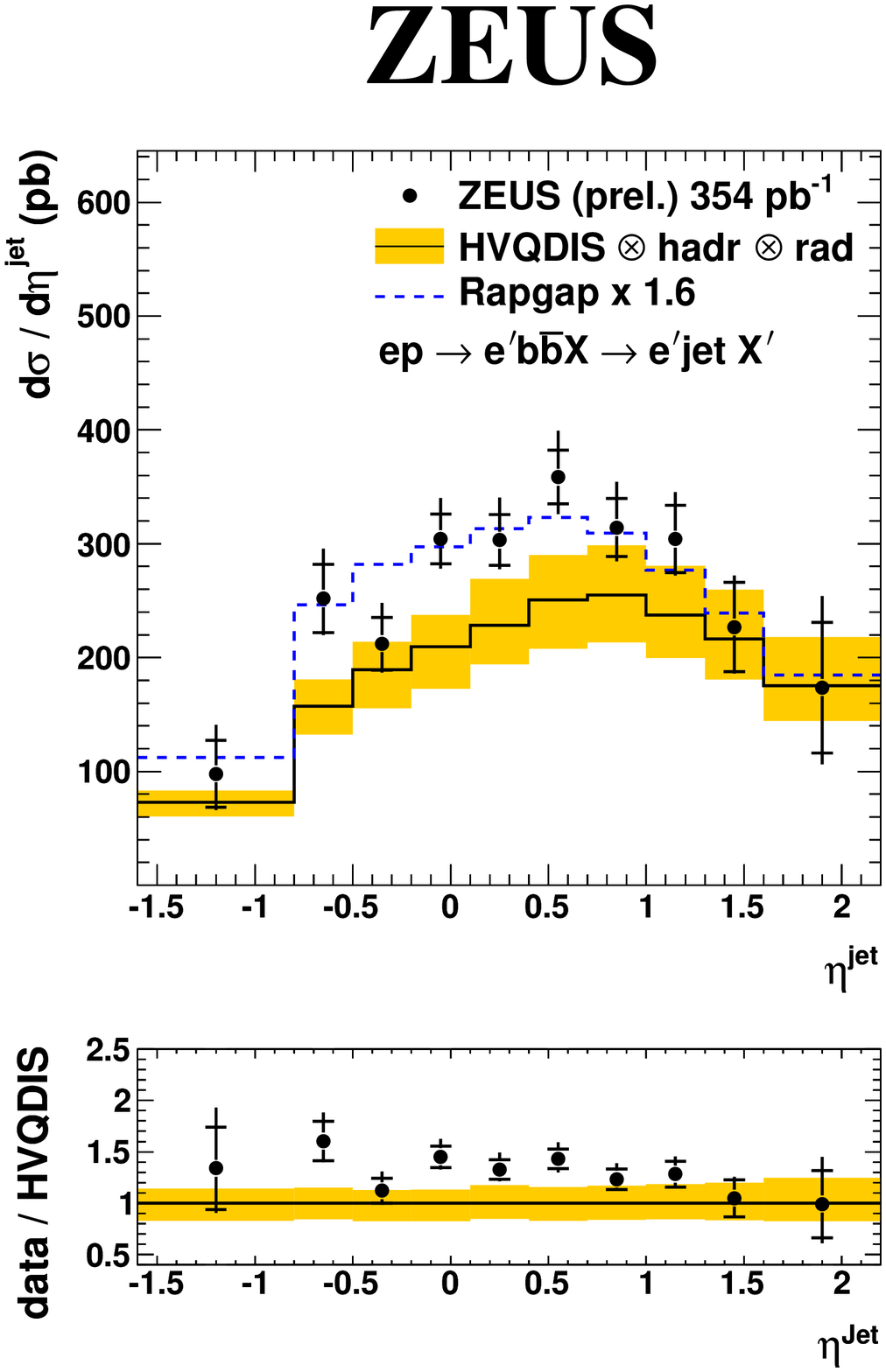}
\end{minipage}
\begin{minipage}[t]{0.24\textwidth}
\centering
\includegraphics[width=0.9\textwidth]{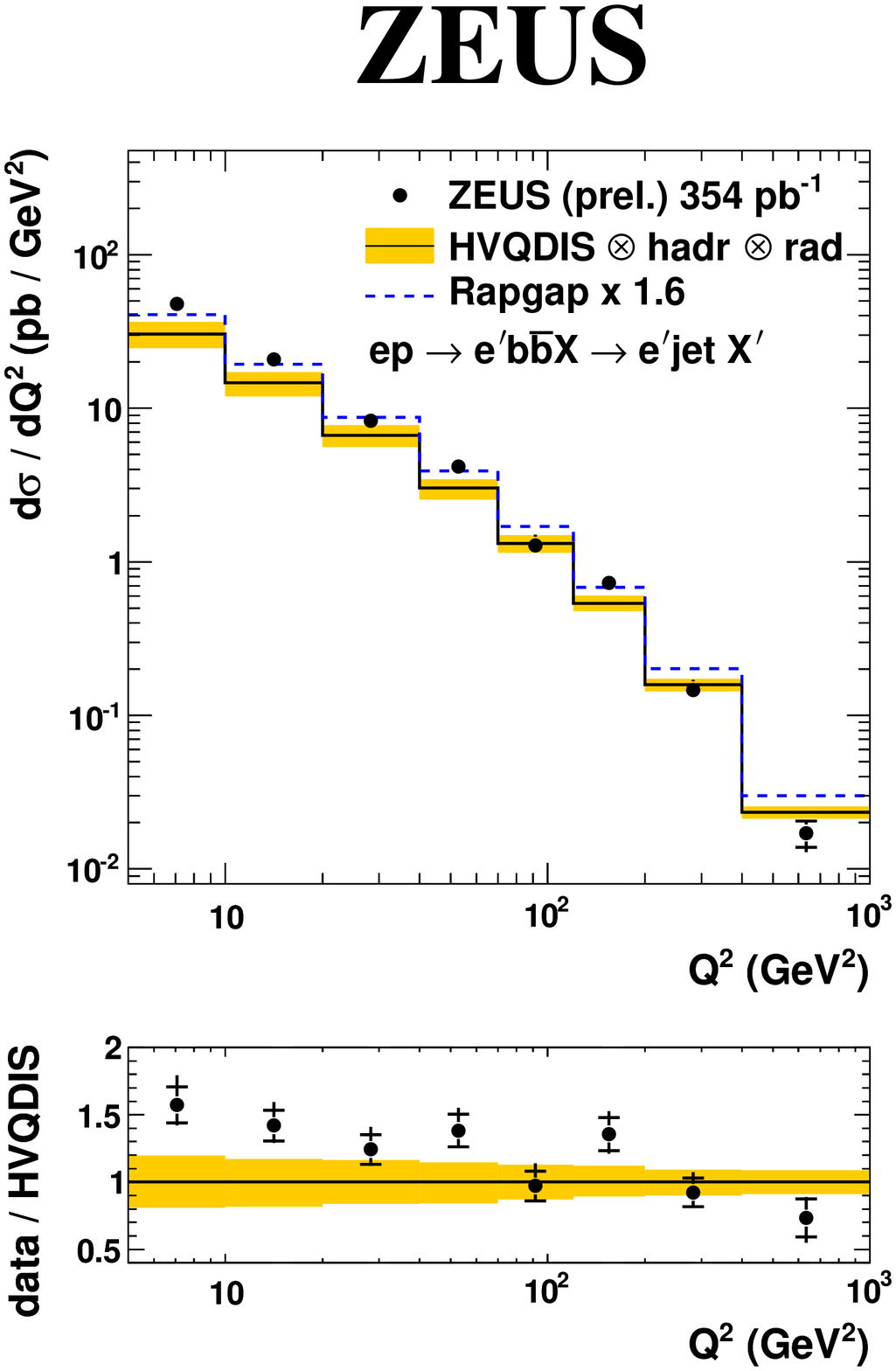}
\end{minipage}
\begin{minipage}[t]{0.24\textwidth}
\centering
\includegraphics[width=0.9\textwidth]{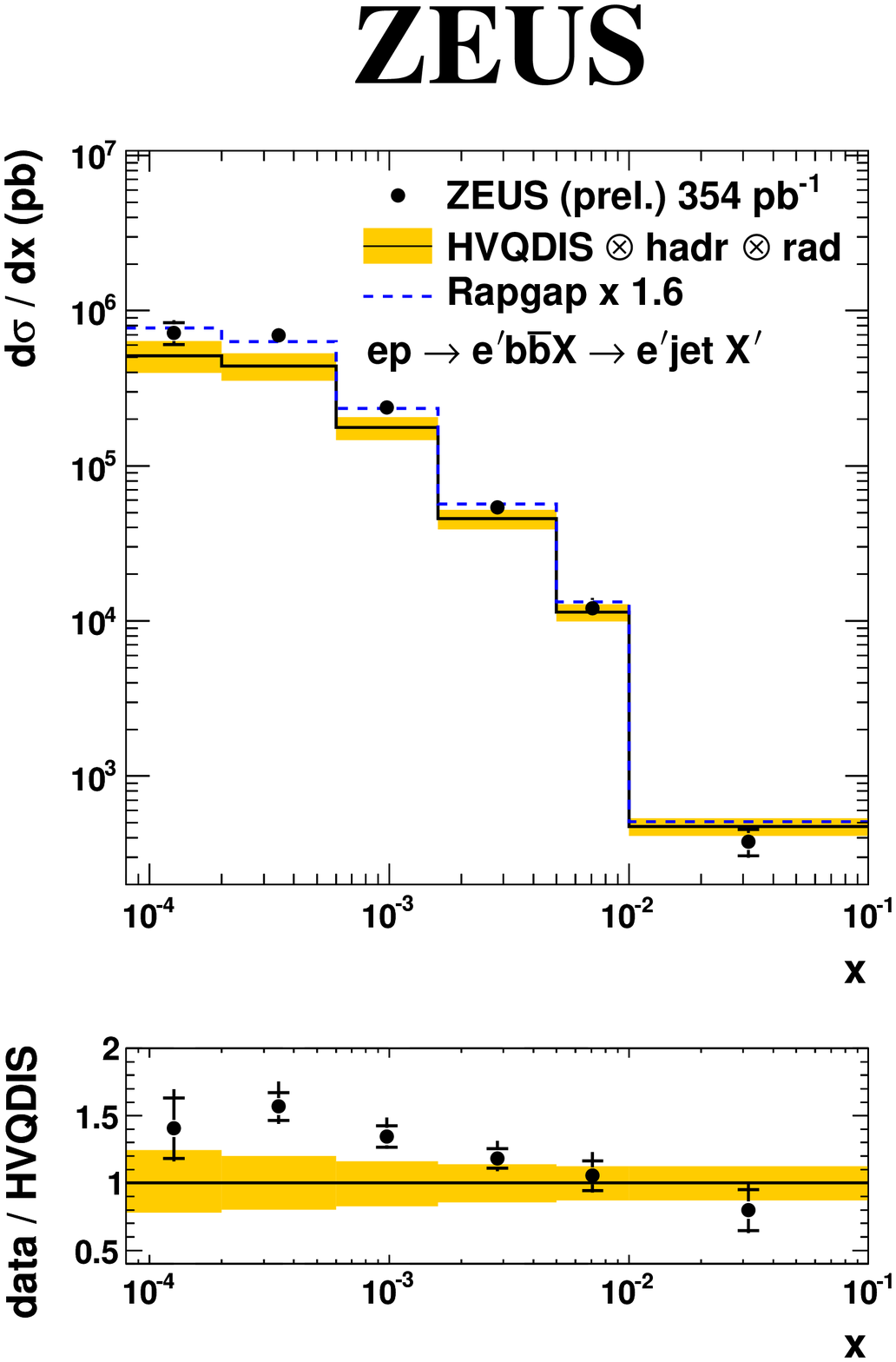}
\end{minipage}
\caption{Differential jet cross sections in beauty events as a function of $E_{T}^{{\rm jet}}$, $\eta^{{\rm jet}}$, $Q^{2}$ and $x$ compared to the NLO QCD calculation of HVQDIS and to the RAPGAP MC.}
\label{fig:jet_cross_sections}
\end{figure}

To extract the contributions from beauty, charm, and LF in the data sample, a binned $\chi^{2}$ fit of the mirrored significance distributions simultaneously in all three mass bins was performed. The overall normalisation of the MC was constrained to the normalisation of the data in the unmirrored significance distributions. The same procedure was followed for every bin of a given observable to obtain differential cross sections.

Differential cross sections for jet production in beauty events as a function of $E_{T}^{{\rm jet}}$, $\eta^{{\rm jet}}$, $Q^{2}$ and $x$ compared to the predictions obtained from HVQDIS and the RAPGAP MC scaled to the data are shown in Fig.~\ref{fig:jet_cross_sections}. The shapes of the measured cross sections are described by HVQDIS and the RAPGAP MC. Typically the data are 20-30\% above the HVQDIS NLO prediction. Double differential cross sections in $x$ and $Q^{2}$ have been extracted. Hence the extraction of $F_{2}^{b\bar{b}}$ from this analysis is imminent.

\section{Conclusion}

Beauty production has been measured using events with a jet and a muon and using inclusive secondary vertices. Single differential cross sections are in reasonable agreement with NLO QCD predictions from HVQDIS. $F_{2}^{b\bar{b}}$ was extracted from the jet+muon measurement and is in good agreement with previous measurements. The precision of $F_{2}^{b\bar{b}}$ will improve using the secondary vertex measurement.


\begin{thebibliography}{99}

\bibitem{ref:muon_and_jet} ZEUS Coll., H.~Abramowicz et al., DESY-10-047, arXiv:1005.3396 [hep-ex] (2010).

\bibitem{ref:hvqdis} B.W.~Harris and J.~Smith, Nucl. Phys. {\bf B 452}, 109 (1995); \\
B.W.~Harris and J.~Smith, Phys. Lett. {\bf B 353}, 535 (1995). Erratum-ibid {\bf B 359} 423 (1995); \\
B.W.~Harris and J.~Smith, Phys. Rev. {\bf D 57}, 2806 (1998).

\bibitem{ref:rapgap} H.~Jung, Comp. Phys. Comm. {\bf 86}, 147 (1995).

\bibitem{ref:ariadne} L.~L\"{o}nnblad, Comp. Phys. Comm. {\bf 71}, 15 (1992).

\bibitem{ref:zeus_2005} ZEUS Coll., S.~Chekanov et al., Eur. Phys. J. {\bf C 65}, 65 (2009).

\bibitem{ref:h1} H1 Coll., A.~Aktas et al., Eur. Phys. J. {\bf C 40}, 349 (2005); \\
H1 Coll., A.~Aktas et al., Eur. Phys. J. {\bf C 45} 23 (2006); \\
H1 Coll., A.~Aktas et al., Eur. Phys. J. {\bf C 65} 89 (2009).

\bibitem{ref:predictions} CTEQ Coll., P.M.~Nadolsky et al., Phys. Rev. {\bf D 78}, 013004 (2008); \\
A.D.~Martin et al., Eur. Phys. J. {\bf C 63}, 189 (2009); \\
M.~Gl\"{u}ck, P.~Jimenez-Delgado and E.~Reya, Eur. Phys. J. {\bf C 53}, 355 (2008); \\
S.~Alekhin et al., Phys. Rev. {\bf D 81}, 014032 (2010); \\
S.~Alekhin and S.~Moch, Phys. Lett. {\bf B 672}, 166 (2009).

\end{thebibliography}
\end{document}